\documentstyle[aps,multicol]{revtex}
\begin{document}

\title{Memory function approach to the Hall constant in strongly 
       correlated electron systems. Part II}

\author{Ekkehard Lange}

\address{Institut f\"ur Theorie der Kondensierten Materie,
Universit\"{a}t Karlsruhe, 76128 Karlsruhe,\\ Germany\\}


\author{\parbox[t]{5.5in}{\small The anomalous frequency and doping
dependence of the Hall constant in the normal state of high-$T_c$ 
superconductors are investigated within models of strongly correlated 
electron systems. In Mori theory, the transition of the Hall constant
from infinite to zero frequency is described by a memory function. It 
naturally introduces a second time scale, that, within the $t-J$
model, is identified with the spinon relaxation time of Anderson. This
provides us with a phenomenological understanding of the interplay
between the frequency and temperature dependence of the Hall constant
for frequencies below the Mott-Hubbard gap. Within the single-band 
Hubbard model in the limit $U\gg t$ ($U$: Coulomb repulsion, $t$: 
hopping amplitude), the memory function is calculated via its moments 
and shown to project out the high-energy scale $U$, introduced by
doubly occupied sites. This causes the Hall constant to decrease by a 
factor $(1+\delta)/2$ ($\delta$: doping), when the frequency is
lowered from infinity to values within the Mott-Hubbard gap. 
Finally, it is outlined, how the Hall constant may be calculated in
the low frequency regime.}} 


\maketitle

\normalsize

\begin{multicols}{2}

\section{Introduction}

Recently, we have proposed a memory function treatment of the Hall
effect in strongly correlated electron systems \cite{lange}. Its main 
advantage consists in the fact that we are directly provided with a
representation of the Hall resistivity. Therefore, in contrast to
ordinary approaches, we may dodge the issue of coping with a quotient 
of conductivities, which may be precarious due to resonances like the
Drude peak. 

In our approach, the frequency dependent Hall constant is given as the
sum of its infinite frequency limit and a memory function
contribution. A perturbative calculation of both terms within the
single-band Hubbard model has demonstrated the usefulness of this
approach \cite{lange}. Our results turned out to be in qualitative
agreement with the unusual experimental findings for high-$T_c$ 
superconductors as, e.g., La$_{2-\delta}$Sr$_{\delta}$CuO$_4$, 
except for the doping dependence of the Hall constant in the vicinity 
of half filling, i.e.\ below $\delta\sim0.1$. There, a $1/\delta$ law 
is observed which requires a description in the strong-correlation 
regime \cite{ong1}. This parameter regime will be the subject of the 
present paper, which is organized as follows. 

In Sec.\ II, the main results concerning the representation of the
Hall constant within Mori theory are compiled from Ref.\ \cite{lange}.
Then, in Sec.\ III, we study the moments of the memory function in the
limit $U\rightarrow\infty$. Thereby, the memory function is found to 
eliminate the high-energy scale set by the Hubbard repulsion $U$. 
This provides us with an explanation of the frequency dependence of
the Hall constant in the cross-over regime from $\omega\gg U$ to
$W\ll\omega\ll U$, where $W$ is the bare band width. Furthermore, a 
simple analytical treatment of the Hall effect within the Hubbard I 
approximation is presented, the results of which turn out to be in
full accord with that obtained by high-temperature expansions. In
Sec.\ IV, we reformulate the frequency dependent Hall constant within
the $t-J$ model in order to address the low frequency regime as
well. The comparison to a representation of $R_H(\omega)$ in terms of
two relaxation rates and effective masses introduced by Anderson
\cite{anderson} provides us with an interpretation of the additional 
relaxation rate $\tau_H$ in the language of the Mori theory and a 
phenomenological understanding of the anomalous frequency and 
temperature dependence of the Hall constant in high-$T_c$ 
superconductors. Finally, we explain how the emerging picture may be
put onto the basis of a microscopic calculation. In this context, we 
reduce the problem of calculating the memory function in the $t-J$
model to the much easier one of finding the first few moments of the 
ordinary current-current correlation functions. We conclude this paper
in Sec.\ V with a discussion of our main results.

\section{Memory function approach to the Hall constant}

For a more detailed introduction into the following memory function 
treatment of the Hall effect in strongly correlated electron systems, 
the reader is referred to Ref.\ \cite{lange}.

We choose the simplest model of strongly correlated
electrons, namely the single-band Hubbard model on a two-dimensional
square lattice:
\begin{equation}
   \hat{H}=-t\sum_{<ij>\sigma} P_{ij}c_{i\sigma}^+c_{j\sigma}
           +U\sum_i \hat{n}_{i\uparrow}\hat{n}_{i\downarrow}\;.
\label{hubbard}
\end{equation}
The second term is a local repulsive interaction. The sum in the 
hopping term is restricted to nearest neighbors and the Peierls phase 
factor $P_{ij}\simeq\exp\{ie\vec{A}(\vec{R}_j)(\vec{R}_i-\vec{R}_j)\}$
guarantees the gauge invariance \cite{peierls} (the sign of the charge
$e$ is chosen to be negative). The vector potential decomposes into 
two terms describing the electric and magnetic field, respectively: 
$\vec{A}(t,\vec{r}\,)=\vec{A}^{el}(t)+\vec{A}^{mag}(\vec{r}\,)$, 
$\vec{E}(t)=-\frac{\partial}{\partial t}\vec{A}^{el}(t)$, and
$\vec{H}=rot\,\vec{A}^{mag}(\vec{r}\,)$. In linear response theory,
the current operators are defined as: 
\begin{equation}
 \hat{J}_{\nu}:=-\frac{1}{e}\,\frac{\delta\hat{H}(t)}{\delta 
 A_{\nu}^{el}(t)}\Bigg|_{A^{el}=0\,.}
\end{equation}
A suitable representation of the Hall constant is obtained within
Mori theory. This formalism enables us to separate the dynamics of the
current operators $\hat{J}_x$ and $\hat{J}_y$ from that of all the
other degrees of freedom, the influence of which is then accumulated
in memory functions \cite{fick}. In the center of this separation
stand the superprojectors 
\begin{eqnarray}
 P&=&(\beta/\chi^0)\sum_{\nu=x,y}|\hat{J}_{\nu})(\hat{J}_{\nu}|\,,
\label{projectorP}\\
 Q&=&1-P\,,
\label{projectorQ}
\end{eqnarray}
acting in operator space ${\cal L}$ and being defined with respect to 
Mori's scalar product 
\begin{equation}
 (\hat{A}|\hat{B}):= (1/\beta)\int_0^{\beta}d\tau\,
 <\exp\{\tau L\}\hat{A}^+\cdot\hat{B}>\,.
\end{equation}
Here, $\beta$ is the inverse temperature, the Liouville operator $L$ 
maps a given operator onto its commutator with the Hamiltonian: 
$L\hat{A}=[\hat{H},\hat{A}]$, and $\chi^0=\beta(\hat{J}_x|\hat{J}_x)$
is the static current susceptibility. This scalar product implies the
so-called Kubo identity
\begin{equation}
 \beta(\hat{A}|L|\hat{B})=<[\hat{A}^+,\hat{B}]>\;,
\label{kubo}
\end{equation}
which will play an important role in the following sections. By 
representing the current-current correlation function within the Kubo 
formula for the conductivity tensor in terms of a memory matrix, we 
obtain the following representation of the Hall constant:  
\begin{equation}
 R_H(z)=\frac{N}{ie^2\chi^0}\lim_{H\rightarrow0}\frac{
        \Omega+iM(z)}{H}\;.
\label{hall}
\end{equation}
The first term is given by the cyclotron frequency 
$\Omega=(1/\chi^0)<[\hat{J}_x\,,\,\hat{J}_y]>$ while the second is
a memory function contribution:
\begin{equation}
 M(z)=\frac{\beta}{\chi^0}(QL\hat{J}_x|
                    \frac{i}{z+QLQ}|QL\hat{J}_y)\,.
\label{memory}
\end{equation}
Here, $z$ is a complex frequency that ultimately has to be continued 
analytically to $z=\omega+i0^+$. This function has the structure of a
relaxation function for the so-called residual forces 
\begin{equation}
 \hat{f}_{\nu}=iQL\hat{J}_{\nu}\,,
\label{residualforce} 
\end{equation}
whose dynamics is governed by the projected Liouville operator $QLQ$
rather than $L$ \cite{fick}. Thus, these forces may vary on a time
scale that is different from that of the current operators
$\hat{J}_{\nu}$. In Sec.\ \ref{phenomenology}, we shall identify these
two time scales with the relaxation rates of the holon and spinon
degrees of freedom in Anderson's tomographic Luttinger liquid theory 
\cite{anderson}. The memory function (\ref{memory}) can be represented
as a spectral integral
\begin{equation}
 M(z)=\int\frac{d\omega}{\pi}\,\frac{M''(\omega)}{\omega-z}\,,
\end{equation} 
where the spectral function $M''(\omega)$ is given by the
discontinuity across the real axis:
\begin{equation}
 M(\omega\pm i0^+)=M'(\omega)\pm iM''(\omega)\,,
\label{spectral}
\end{equation} 
and can be shown to be real and even. Since the memory function
vanishes as $1/z^2$ at high frequencies, the first term on the Rhs.\
of Eq.\ (\ref{hall}) represents the Hall constant in the limit 
$\omega\rightarrow\infty$, that was considered by Shastry et al. 
\cite{SSS}. It will be denoted as $R_H^{\infty}$ from now on.
 
\section{Hall constant in the cross-over regime from $\omega\gg U$ 
         to $\omega\ll U$}    
  
Numerical studies of multi-band Hubbard models for the Cu-O planes of
high-$T_c$ materials indicate that all low-energy excitations are 
reproducible within a single-band Hubbard model with $U\sim W$
\cite{bacci}. Unfortunately, this parameter regime is not accessible
to reliable analytical calculations. Therefore, we shall exaggerate
the impact of $U$ by considering the range $U\gg W$ instead.

\subsection{The moments of the memory function in the limit 
               $U\rightarrow\infty$} 

The main problem in dealing with the memory function (\ref{memory}) is
related to the fact that its dynamics is governed by the projected
Liouville operator $QLQ$ rather than $L$. We resolve this difficulty
by inquiring into the properties of the memory function via its
moments. Then, we may get rid of the superprojectors $Q$ simply by 
resorting to their definition through Eqs.\ (\ref{projectorP}) and 
(\ref{projectorQ}).

We start by writing the Hall constant (\ref{hall}) in terms of
spectral functions:
\begin{equation}
 R_H(z)=\frac{N}{e^2(\chi^0)^2H}\,\int_{-\infty}^{\infty}d\omega\,
        \left\{S_H(\omega)+K(\omega)\,\frac{\omega}{\omega-z}\right\}.
\label{frha}
\end{equation}
The first spectral function in the integrand is defined as 
\begin{equation}
 S_H(\omega)=-i<[\hat{J}_x,\delta(\omega+L)\hat{J}_y]>\,,
\label{SHofomega}
\end{equation}
and corresponds to the Hall matrix element of the current-current 
correlation function. The second function is related to the spectral
function $M''(\omega)$, introduced in Eq.\ (\ref{spectral}), by
$K(\omega)=\chi^0M''(\omega)/(\pi\omega)$, which implies:
\begin{equation}
 K(\omega)=\frac{-i\beta}{\omega}(QL\hat{J}_x|\delta(\omega+QLQ)
                                                  |QL\hat{J}_y)\;.
\label{Kofomega}
\end{equation}
This function describes the correlation between the residual forces
$\hat{f}_x$ and $\hat{f}_y$. In the following, we shall unravel a 
simple connection between the moments of the two functions
$S_H(\omega)$ and $K(\omega)$ in the limit $U\rightarrow\infty$. 

First, we note that both functions may be shown to be even and real
\cite{lange,fick}. Furthermore, we expect both functions to vanish 
beyond a certain frequency above $U$, since $U$ is the highest energy 
scale in the problem. This assumption will be corroborated below up 
to mistakes of the order $t/U$. Hence, for finite but large $U$, all 
moments exist and it is sufficient, to consider only the even ones:
\begin{eqnarray}
 S_l&=&\int_{-\infty}^{\infty}d\omega\,S_H(\omega)\omega^{2l},
\label{Smomdef}\\
 K_l&=&\int_{-\infty}^{\infty}d\omega\,K(\omega)\omega^{2l},
\label{Kmomdef}
\end{eqnarray}
where $l\ge0$. Since we are mainly interested in the dc-Hall constant,
we would like to calculate $K_0$. Unfortunately, this is not feasible 
on the basis of Eq.\ (\ref{Kofomega}), since the inverse of the 
projected Liouville operator $QLQ$ does not exist. However, all the 
other moments (\ref{Kmomdef}) can be calculated: Due to Eq.\ 
(\ref{Kofomega}), they are given as: 
\begin{equation}
 K_l=i\beta(\hat{J}_x|\underbrace{LQL\;\ldots\;LQL}_{\mbox{$2l$
        projectors $Q$}}|\hat{J}_y)\quad(l\ge1)\;.
\label{Kmom}
\end{equation}
Now, if we insert $Q=1-P$, this expression decomposes into $4^l$
terms. Consider a special one consisting of $p$ superprojectors $P$. 
By using the definition of $P$, this term is seen to decompose further
into $f=p+1$ factors of the form $\beta(\hat{J}|L^{m_j}|\hat{J})$, the
orders of which are $U^{m_j-1}$ due to the Kubo identity (\ref{kubo}).
Here, $\sum_{j=1}^f m_j=2l+1$. Thus, each factor lowers the relative 
order in $U$ by one. Hence, the more superprojectors $P$ a given term 
is composed of, the lower its relative order in $U$ is. Therefore, all 
superprojectors $Q$ may be removed from the Rhs.\ of Eq.\ (\ref{Kmom})
to leading order in $t/U$. And this, in turn, establishes the
following relation between the $l\ge1$ moments of Eqs.\
(\ref{Smomdef}) and (\ref{Kmomdef}):
\begin{equation}
 \frac{K_l}{U^{2l}}=-\,\frac{S_l}{U^{2l}}+o(\frac{t}{U}).
\label{momeq}
\end{equation}
This equation does {\em not} imply that the functions
(\ref{SHofomega}) and (\ref{Kofomega}) differ only by a sign in the
limit $U\rightarrow\infty$. This conclusion would require positive or
negative definite functions and finite moments even in the limit 
$U\rightarrow\infty$. None of both conditions is satisfied. To proceed
anyway, we remind ourselves that in the context of the Hubbard model
in the strong correlation limit, any spectral function is believed to
separate into individual ``peaks'' centered around integer multiples
of $U$ \cite{harrislange,donald,eskes} (in this context, any connected 
structure of a given spectral function, irrespective of its detailed
shape, is referred to as a ``peak''; for instance, it may vanish at 
discrete points). This reflects the fact that one-particle excitations
may be grouped into two Hubbard bands separated by the so-called
charge transfer gap, which is of the order $U$. Since the current 
operators produce particle-hole excitations, we expect the functions 
(\ref{SHofomega}) and (\ref{Kofomega}) to have peak structures
centered around $\omega=0$ and $\pm U$ related to excitations within
the two Hubbard bands and across the charge transfer gap,
respectively. Therefore, these peaks are expected to have widths of
the order of those of the Hubbard bands. In the following, we shall
prove this picture at least for the function (\ref{SHofomega}) and
derive formulas required to extract more information from the
relations (\ref{momeq}). 

\subsection{Structure of the functions $S_H(\omega)$ and
$K(\omega)$ \label{structureofSK}}

The appropriate technique to investigate spectral properties of the
Hubbard model in the strong correlation limit was pioneered by Harris
and Lange in the special case of single-particle  excitations 
\cite{harrislange} and generalized by several other authors, see for
instance Refs.\ \cite{donald,eskes}. At the heart of this procedure
stands the decomposition of a given operator into terms, which increase 
the number of doubly occupied sites by integer values $p$:
\begin{equation}
 \hat{O}=\sum_{p=-\infty}^{\infty}\hat{O}_{pU}\,.
\label{decompoperator}
\end{equation}
Together with the Lehmann representation of a given spectral function,
one may then address the properties of its individual peaks. The 
decomposition (\ref{decompoperator}) is accomplished by an iterative 
procedure based on a canonical transformation of the Hubbard
Hamiltonian: 
$\hat{H}\rightarrow\exp\{i\hat{S}\}\hat{H}\exp\{-i\hat{S}\}$. The 
expansion of the operator $\hat{S}$ up to the order $l$ in $t/U$
eliminates those processes from $\hat{H}$ which change the total
number of doubly occupied sites up to the order $t^l/U^{l-1}$. The 
corresponding transformed Hamiltonian $\hat{H}^{(l+1)}$ in turn helps 
to fix the next order of $\hat{S}$ in $t/U$ and so on. Thus,
subsequent iterations generate increasing orders in $t/U$. Once the 
generator $\hat{S}$ has been found to a given order, one may decompose
any operator to the same order by first decomposing its rotated
counterpart \cite{eskes}.  

In our case, $t/U$-expansions may be terminated after the zeroth order
term since the relation (\ref{momeq}) indicates that it is not
sensible to go beyond. Then, we do not have to distinguish between
original and transformed Fermi operators and the decomposition 
(\ref{decompoperator}) specialized to the case of the operators 
$\hat{D}_{ij}^{\sigma}\equiv c_{i\sigma}^+c_{j\sigma}$, making up 
the components of the current operator, becomes:
\begin{equation}
 \hat{D}_{ij}^{\sigma}=\hat{D}_{ij;-U}^{\sigma}
                      +\hat{D}_{ij;0}^{\sigma}
                      +\hat{D}_{ij;U}^{\sigma}\,.
\label{expa0}
\end{equation}
In terms of Hubbard operators 
$X_i^{0\sigma}\equiv c_{i\sigma}(1-\hat{n}_{i\bar{\sigma}})$,
$X_i^{\sigma0}\equiv (1-\hat{n}_{i\bar{\sigma}})c_{i\sigma}^+$,
$X_i^{\bar{\sigma}2}\equiv c_{i\sigma}\hat{n}_{i\bar{\sigma}}$,
and $X_i^{2\bar{\sigma}}\equiv \hat{n}_{i\bar{\sigma}}c_{i\sigma}^+$,
where $\bar{\sigma}\equiv-\sigma$, the terms of Eq.\ (\ref{expa0}) 
may be written conveniently as:
\begin{eqnarray}
 \hat{D}_{ij;0}^{\sigma}&=&X_i^{\sigma0}X_j^{0\sigma}
                        +X_i^{2\bar{\sigma}}X_j^{\bar{\sigma}2}\,,
\label{Dij0}\\
 \hat{D}_{ij;-U}^{\sigma}&=&X_i^{\sigma0}X_j^{\bar{\sigma}2}\,,\\
 \hat{D}_{ij;U}^{\sigma}&=&X_i^{2\bar{\sigma}}X_j^{0\sigma}\,.
\end{eqnarray}
The Lehmann representation of the function (\ref{SHofomega}) is
derived straightforwardly:
\begin{eqnarray}
 S_H(\omega)&=&\frac{1}{Z}\sum_{n,m}M_{nm}
             (e^{-\beta\epsilon_n}-e^{-\beta\epsilon_m})\nonumber\\
       &&\times\delta(\omega-[\epsilon_n-\epsilon_m])\,,\\
 M_{nm}&=&\frac{1}{2i}\{<n|\hat{J}_x|m><m|\hat{J}_y|n>\nonumber\\
       &&              -<n|\hat{J}_y|m><m|\hat{J}_x|n>\}\,, 
\end{eqnarray}
where states and energies are defined through the eigenvalue equation 
$(\hat{H}-\mu\hat{N})|n>=\epsilon_n|n>$. Inserting the decomposition 
of the current operators corresponding to (\ref{expa0}), we find that
the peak centered around $\omega=pU$ has a weight given by
\begin{equation}
 W_{pU}=\frac{1}{2i}(<[\hat{J}_{x;pU},\hat{J}_{y;-pU}]>
                    +<[\hat{J}_{x;-pU},\hat{J}_{y;pU}]>)
\end{equation}
and that only the peaks $p=0$ and $p=\pm1$ survive in leading order in
$t/U$. 

We assume that the function (\ref{Kofomega}) has qualitatively the
same triple-peak structure. Although not proven, this assumption is
shown to lead to reasonable conclusions.

\subsection{Frequency dependence of the Hall constant in the range 
            $\omega\gg W$}    

Given the peak structure of the functions (\ref{SHofomega}) and 
(\ref{Kofomega}), only the contributions of the satellite peaks around
$\omega=\pm U$ can be resolved in the $l\ge1$ moments (\ref{Smomdef})
and (\ref{Kmomdef}). Thus, the relation (\ref{momeq}) implies that the
``spectral weights'' of the peaks of $K(\omega)$ and $S_H(\omega)$ 
around $\omega=U$ differ only by a sign. Together with Eq.\
(\ref{frha}), we may then draw the following conclusions: For
$\omega\gg U$, all peaks of $S_H(\omega)$ contribute to the Hall 
constant and none of $K(\omega)$. In the frequency range
$W\ll\omega\ll U$, the high-frequency peaks cancel each other out
while the contribution of the zero-frequency peak of $K(\omega)$ is
negligible. Within this charge transfer gap region, the frequency 
dependent Hall constant is then lowered by a factor $p$ in comparison 
to its infinite frequency limit,
\begin{equation}
 R_H^*\equiv R_H(W\ll\omega\ll U)=p\,R_H^{\infty},
\label{result}
\end{equation}
if we define $p$ to be the relative spectral weight of the low energy 
structure of $S_H(\omega)$: 
\begin{equation}
 p=\frac{W_0}{W_{-U}+W_0+W_{U}}=\frac{<[\hat{J}_{x;0},\hat{J}_{y;0}]>}
                                     {<[\hat{J}_x,\hat{J}_y]>}\;.
\label{ppp}
\end{equation}
Eq.\ (\ref{result}) is valid except for mistakes of the order $t/U$.
Therefore, it is sufficient to evaluate it in the limit 
$U\rightarrow\infty$. 

First, we seek for an interpretation of $R_H^*$. We begin by noting
that the infinite frequency Hall constant, i.e.\ the first term of
Eq.\ (\ref{hall}), may be rewritten as follows:
\begin{equation}
 R_H^{\infty}=\frac{1}{e}\frac{2a_d}{(2a_n)^2}\;,
\label{R_Hinf}
\end{equation}
where $a_n$ and $a_d$ is the amplitude of a nearest neighbor hop and a
hop diagonally across the unit cell, respectively. This holds on
account of the following relations, which may be proven by
straightforward analysis: 
\begin{eqnarray}
 <[\hat{J}_x,\hat{J}_y]>&=&8iNet^2Ha_d\,,
\label{commutator}\\
 \chi^0&=&4tNa_n\,.
\end{eqnarray}
In the limit $U\rightarrow\infty$, we have
\begin{eqnarray}
 \lim_{U\rightarrow\infty}a_n
   &=&<X^{\sigma0}_{\vec{R}}X^{0\sigma}_{\vec{R}+\hat{x}}>\,,
\label{a_nUinf}\\
 \lim_{U\rightarrow\infty}a_d
   &=&<X^{\sigma0}_{\vec{R}}X^{0\sigma}_{\vec{R}+\hat{x}+\hat{y}}>\,,
\label{a_dUinf}
\end{eqnarray}
where $\hat{x}$ and $\hat{y}$ is a primitive lattice vector in the
$x$- and $y$-direction, respectively. Here and in the following,
expectation values are taken with respect to states without double 
occupancies. Then, the projected current operators appearing in the 
nominator of the Rhs.\ of Eq.\ (\ref{ppp}) take on the following form:
\begin{equation}
 \hat{J}_{\nu;0}=it\sum_{<ij>\sigma}\Delta_{ij}^{\nu}P_{ij}
                        X_i^{\sigma0}X_j^{0\sigma}\,,
\label{currentUinf}
\end{equation}
since the term $X_i^{2\bar{\sigma}}X_j^{\bar{\sigma}2}$ of Eq.\ 
(\ref{Dij0}) may be omitted. Here, 
$\vec{\Delta}_{ij}\equiv\vec{R}_i-\vec{R}_j$ and $P_{ij}$ is the
phase factor defined in the text following Eq.\ (\ref{hubbard}).
From Eqs.\ (\ref{ppp})-(\ref{currentUinf}), we conclude that $R_H^*$,
defined in Eq.\ (\ref{result}), represents the infinite frequency Hall 
constant of the $U=\infty-$Hubbard model, which is defined as follows:
\begin{equation}
 \hat{H}=-t\sum_{<ij>\sigma}P_{ij}X_i^{\sigma0}X_j^{0\sigma}\,.
\label{U=infty}
\end{equation}
In fact, the analysis of Sec.\ II is straightforwardly carried over to
this model, the current operator of which is then found to be given by
Eq.\ (\ref{currentUinf}). Note, that $p\ne1$ expresses the fact that
the limits $\omega\rightarrow\infty$ and $U\rightarrow\infty$ do not
commute: If we start with the limit $\omega\rightarrow\infty$, the 
integral over $S_H(\omega)$ in Eq.\ (\ref{frha}) extends over all
three peaks while when taking the limits in reversed sequence, the
high-energy peaks are unattainable from the outset. 

Next, we derive an exact analytical expression for $p$ by taking the
additional limit $T\rightarrow\infty$. We may wonder whether this is
reasonable. However, since the limit $U\rightarrow\infty$ was already
carried out, at least, the condition $T\ll U$ is satisfied, i.e., the
thermal energy cannot excite an electron across the charge transfer
gap. Furthermore, we expect neither of the high-frequency objects
$R_H^{\infty}$ and $R_H^*$ to depend appreciably on temperature, since
they correspond to and generalize the semi-classical expression for
the Hall constant \cite{lange}. In the context of high-temperature 
expansions, one has to cope with electrons or holes hopping around 
closed loops, which are defined by a sequence of adjacent lattice 
sites. Therefore, it is convenient not to expand the phase factors. 
Then, an electron hopping along a polygon $ijk\ldots li$ accumulates a
phase proportional to the flux $\phi_{ijk\ldots li}$ enclosed. The 
procedure to expand expectation values of Hubbard operators like that 
of Eqs.\ (\ref{a_nUinf}) and (\ref{a_dUinf}) or the nominator of Eq.\ 
(\ref{ppp}) in powers of $1/T$ is explained, e.g., in Ref.\ 
\cite{thompson}. To leading order, we obtain:
\begin{equation}
 p=\frac{1+\delta}{2}\,,
\label{pppresult}
\end{equation}
where here and in the following, the electron density is measured via
the average number $\delta$ of holes per lattice site introduced into
the half filled system, i.e. $\delta=0$ corresponds to half filling 
while $\delta=1$ corresponds to the empty band. Although the
high-temperature calculation has introduced another high-energy scale 
into our system, we expect this result to hold qualitatively for low 
temperatures as well. For instance, since $p$ is a measure for the 
difference between the plateau values of $R_H(\omega)$ on both sides
of $U$, it is expected to increase monotonically as half filling is 
approached. Further down, we shall derive the same expression for $p$ 
within a simple approximation valid at $T=0$. 

Finally, we calculate $R_H^{\infty}$ to leading order in $1/T$. The 
leading order of the amplitudes (\ref{a_nUinf}) and (\ref{a_dUinf}) 
are found to be $a_n=(\beta t/2)\delta(1-\delta)$ and $a_d=
-(\beta^2 t^2/4)\delta(1-\delta)(1-3\delta)$, which results in
\begin{equation}
 R_H^{\infty}=\frac{1}{|e|}\left(\frac{1-3\delta}{2}\right)
              \left(\frac{1}{\delta}+\frac{1}{1-\delta}\right)\,.
\label{RHinf1}
\end{equation}
Together with Eqs.\ (\ref{result}) and (\ref{pppresult}), we recover
the result for $R_H^*$ of Ref.\ \cite{SSS}, that was derived within
the $t-J$ model in leading order in $1/T$. In this work, it was
further shown that, although $R_H^*$ is renormalized as a function of 
$T$ and $J$ when including higher orders in $1/T$, the doping
dependence of the Hall constant retains its most important features:
its sign change at $\delta\approx1/3$ and its singular behavior in
the vicinity of half filling. What seems to be striking at first sight
is the fact that the same properties are encountered for
$R_H^{\infty}$, i.e.\ the Hall constant at frequencies well beyond
$U$. At such high frequencies, the {\em dynamics} of the electrons is 
insensitive to the interaction $U$. However, for nondynamical
quantities as matrix elements, the correlations introduced by the 
Hubbard interaction remain important. 

Before we discuss possibilities to extend the moments technique to
lower frequencies, we shall rederive the results of this subsection
within a simple approximation, valid at $T=0$.

\subsection{Hubbard I approximation}

An expression for the frequency dependent Hall conductivity with 
vertex corrections having been neglected was derived in Ref.\ 
\cite{voruganti}:
\begin{eqnarray}
 \sigma_{xy}(z)&=&\frac{e^3H}{2}\sum_{\vec{k}\sigma}
    \left[\frac{\partial\epsilon_{\vec{k}}}{\partial k_x}\right]^2
    \frac{\partial^2\epsilon_{\vec{k}}}{\partial k_y^2}\,
    \frac{\Pi_H(z,\vec{k})}{z}\;,
\label{condH}\\
 \Pi_H(i\omega_m,\vec{k})&\equiv&\frac{1}{\beta}\sum_n
    G_{\vec{k}}(i\omega_n)^2\nonumber\\
               &\times&\left[G_{\vec{k}}(i\omega_n+i\omega_m)
                            -G_{\vec{k}}(i\omega_n-i\omega_m)\right].
\label{helpfunct}
\end{eqnarray}
Here, the Green's function is given in terms of its spectral function
as
\begin{equation}
 G_{\vec{k}}(i\omega_n)=\int_{-\infty}^{\infty}\frac{d\omega}{2\pi}\,
                      \frac{A_{\vec{k}}(\omega)}{i\omega_n-\omega}\,, 
\label{green}
\end{equation}
and $e$, in our notation, is negative. Furthermore, it is assumed that
the momentum dependence arises solely from the dispersion
$\epsilon_{\vec{k}}$ of the bare band. In Mori theory, the Hall
conductivity is represented as \cite{lange}:
\begin{equation} 
 \sigma_{xy}(z)=\frac{ie^2}{z}\beta(\hat{J}_x|\frac{L}{z+L}|\hat{J}_y)\,. 
\end{equation}
Therefore, the function (\ref{SHofomega}) may be connected to the
Hall conductivity $\sigma_{xy}(\omega\pm i0^+)\equiv\sigma_H'(\omega)
\pm i\sigma_H''(\omega)$ via
\begin{equation}
 S_H(\omega)=\frac{\omega\sigma_H''(\omega)}{\pi e^2}\,.
\label{SHof2}
\end{equation}
Since $\sigma_H''(\omega)$ arising from Eqs.\ (\ref{condH}) and
(\ref{helpfunct}) may be shown to be real and odd, the function 
(\ref{SHofomega}) has indeed the correct analytic properties.
Inserting Eqs.\ (\ref{helpfunct}) and (\ref{green}) into Eq.\
(\ref{condH}) and using Eq.\ (\ref{SHof2}), we obtain, after a
standard calculation \cite{fetter}:
\begin{eqnarray}
 S_H(\omega)&=&|e|H\sum_{\vec{k}}
    \left[\frac{\partial\epsilon_{\vec{k}}}{\partial k_x}\right]^2
    \frac{\partial^2\epsilon_{\vec{k}}}{\partial k_y^2}\,
    X_{\vec{k}}(\omega)\,,
\label{SHof3}\\
 X_{\vec{k}}(\omega)&\equiv&
            \int_{-\infty}^{\infty}\frac{d\omega_1}{2\pi}
            \int_{-\infty}^{\infty}\frac{d\omega_2}{2\pi}\, 
          A_{\vec{k}}(\omega_1)A_{\vec{k}}(\omega_2)\nonumber\\
 &\times&\left\{\frac{F_{\vec{k}}(\omega_1;\omega)
                        -F_{\vec{k}}(\omega_2;\omega)}
                           {\omega_1-\omega_2}
                     +(\omega\rightarrow-\omega)\right\}\,,
\label{Xabbr}\\
 F_{\vec{k}}(\epsilon;\omega)&\equiv&
             \frac{A_{\vec{k}}(\epsilon-\omega)}{2\pi}
         \left[f(\epsilon)-f(\epsilon-\omega)\right]\,.
\label{Fabbr}
\end{eqnarray}
By means of a partial integration, the corresponding sum rule
is straightforwardly shown to be satisfied (cf.\ Eqs.\
(\ref{SHofomega}), (\ref{commutator}), and (\ref{a_d})).   

Next, we evaluate Eqs.\ (\ref{SHof3})-(\ref{Fabbr}) in the so-called 
Hubbard I approximation \cite{hubIapprox}:
\begin{eqnarray}
 \frac{A_{\vec{k}}(\omega)}{2\pi}
  &=&\frac{1+\delta}{2}\;
     \delta(\omega-\frac{1+\delta}{2}\epsilon_{\vec{k}})\nonumber\\
  &+&\frac{1-\delta}{2}\;
     \delta(\omega-U-\frac{1-\delta}{2}\epsilon_{\vec{k}})\,.
\label{hubbardI}
\end{eqnarray}
At $U=\infty$, only the first term contributes. Then, the quantity 
(\ref{Xabbr}) becomes
\begin{equation}
 X_{\vec{k}}(\omega)=2\left(\frac{1+\delta}{2}\right)^2
    \frac{\partial f(\frac{1+\delta}{2}\,\epsilon_{\vec{k}})}
         {\partial\epsilon_{\vec{k}}}\,\delta(\omega)\,,
\end{equation}
and by means of an integration by parts, we obtain
\begin{equation}
 S_H(\omega;U=\infty)=\frac{1+\delta}{2}\,
        \left(-i<[\hat{J}_x,\hat{J}_y]>\right)\delta(\omega)\,.
\label{HIresult1}
\end{equation}
Here, the expectation value is given by Eqs.\ (\ref{commutator}) and 
(\ref{a_d}). The sum rule obeyed by $S_H(\omega)$ for $U\ne\infty$
implies $p=(1+\delta)/2$, as in Eq.\ (\ref{pppresult}). This provides 
further evidence that the result (\ref{pppresult}) may be trusted for 
all temperatures. Note, that in a real system, the function 
$S_H(\omega)$ must vanish at $\omega=0$. Only then do we obtain a
finite Hall conductivity at $\omega=0$. The result (\ref{HIresult1}) 
is an artifact of the vanishing width of the lower Hubbard band in 
(\ref{hubbardI}). If we consider finite but large values of $U$, we
may simulate the high-energy peaks of the function (\ref{SHofomega})
by delta functions as well:
\begin{eqnarray}
 &&S_H(\omega;U\rightarrow\infty)=-i<[\hat{J}_x,\hat{J}_y]>\nonumber\\
 &&\quad\times\left(\frac{1+\delta}{2}\,\delta(\omega)
      +\frac{1-\delta}{4}\,[\delta(\omega-U)+\delta(\omega+U)]\right)\,.
\label{HIresult2}
\end{eqnarray}
This may be proven by calculating the moments (\ref{Smomdef}) in the
limit $U\rightarrow\infty$.

Finally, we calculate $R_H^{\infty}$, given by Eq.\ (\ref{R_Hinf}), 
analytically within the approximation (\ref{hubbardI}) and at
$U=\infty$. Although this has been done numerically some time ago 
\cite{fukcomment}, our simple analytical treatment allows for a direct
comparison with the exact high-temperature result (\ref{RHinf1}) and 
demonstrates that the resulting doping dependence does not rely on the
location of the Fermi surface. This last point is blurred in the 
Boltzmann equation based approach of Ref.\ \cite{fukuyama}.

The amplitudes on the Rhs.\ of Eq.\ (\ref{R_Hinf}) may be written as
follows:
\begin{eqnarray}
 a_n&=&\frac{1}{N}\sum_{\vec{k}}\cos k_x n_{\vec{k}\sigma}\,,
\label{a_n}\\
 a_d&=&\frac{1}{N}\sum_{\vec{k}}\cos k_x\cos k_y n_{\vec{k}\sigma}\,.
\label{a_d}
\end{eqnarray}
Here, the density is given as
\begin{equation}
 n_{\vec{k}\sigma}=\frac{1+\delta}{2}\;
   f(\frac{1+\delta}{2}\;\epsilon_{\vec{k}}),
\label{density}
\end{equation}
since in Hubbard I approximation at $U=\infty$, only the first term on
the Rhs.\ of Eq.\ (\ref{hubbardI}) is to be kept. Momentum dependences
arise solely from the dispersion $\epsilon_{\vec{k}}$ of the bare
band, why next, we are looking for expressions for the following 
functions:
\begin{eqnarray}
 A(\epsilon)&\equiv&\frac{1}{N}\sum_{\vec{k}} 
             \cos k_x\,\delta(\epsilon-\epsilon_{\vec{k}})\;,\\
 B(\epsilon)&\equiv&\frac{1}{N}\sum_{\vec{k}} 
             \cos k_x\cos k_y\,\delta(\epsilon-\epsilon_{\vec{k}})\;.
\label{B_edef}
\end{eqnarray}
A convenient way to find smooth approximations for these functions is 
to calculate them in the limit of infinite dimensions $d$ \cite{lange}:
\begin{eqnarray}
 A(\epsilon)&=&-\frac{\epsilon}{\sqrt{d}}\,D^0(\epsilon)\,,
\label{A_e}\\
 B(\epsilon)&=&\frac{1}{d}(\epsilon^2-\frac{1}{2})\,D^0(\epsilon)\,,
\label{B_e}
\end{eqnarray}
where $D^0(\epsilon)=(1/\sqrt{\pi})\exp(-\epsilon^2)$ is the density
of states of the bare system. Except for the prefactors $1/\sqrt{d}$
and $1/d$, which ultimately cancel each other out, the functions
(\ref{A_e}) and (\ref{B_e}) may be compared to those calculated
numerically on a two dimensional lattice. This reveals that the
main effect of the limit $d\rightarrow\infty$ is to smooth out the
logarithmic singularities at zero energy encountered in the case of
the functions $D^0(\epsilon)$ and (\ref{B_edef}) in $d=2$. 
Therefore, this limiting procedure does certainly not affect the
validity of our present analysis in any serious manner. At $T=0$,
Eqs.\  (\ref{a_n})-(\ref{B_e}) imply:
\begin{eqnarray}
 2a_n&=&\frac{1+\delta}{2\sqrt{d}}\,D^0(\frac{2\epsilon_F}{1+\delta})\,,\\
 2a_d&=&\frac{-\epsilon_F}{d}\,D^0(\frac{2\epsilon_F}{1+\delta})\,.
\label{adad}
\end{eqnarray}
A relation between the Fermi energy and the doping parameter is
established straightforwardly, which, in terms of the function
\begin{equation}
 H(\delta)\equiv\frac{2\epsilon_F(\delta)}{1+\delta}\,,
\label{Hdef}
\end{equation}
may be written as: 
\begin{equation}
 \frac{1-3\delta}{1+\delta}=\mbox{erf}(H(\delta))\,.
\label{Hres}
\end{equation}
In summary, the function (\ref{R_Hinf}) at $U=\infty$ and $T=0$ may be
written as follows:
\begin{equation}
 |e|R_H^{\infty}=\frac{2}{1+\delta}\,\frac{H(\delta)}{D^0(H(\delta))}\,.
\label{RHinf2}
\end{equation}
The most important features are: Firstly, at $\delta=1/3$,
$R_H^{\infty}$ vanishes. Secondly, at $\delta\rightarrow1$, we recover
the exact result $R_H^{\infty}\simeq-1/(|e|(1-\delta))$. And thirdly,
at $\delta\rightarrow0$, we find $R_H^{\infty}\simeq1/(|e|2\delta)$. 
The last two statements are proven with the asymptotic relation 
$D^0(\epsilon)/\epsilon\simeq\pm1-\mbox{erf}(\epsilon)$, valid in the 
limit $\epsilon\rightarrow\pm\infty$. All these points are in exact 
agreement with (\ref{RHinf1}). We take this as an indication that, on
the one hand, the high-temperature result (\ref{RHinf1}) remains 
qualitatively valid even at low temperatures, and, on the other, that
the Hubbard I approximation is remarkably good in the case of the 
quantity (\ref{R_Hinf}). In Ref.\ \cite{fukuyama}, the doping
dependence of the Hall constant in Hubbard I approximation was
discussed in terms of the Fermi surface. This is misleading for two 
reasons. 

For one thing, the Hubbard I approximation misplaces the Fermi
surface: The Luttinger theorem, which relates the volume enclosed by
the Fermi surface to the electron density \cite{agd}, is violated in
this approximation. In contrast to this, angle resolved photoemission 
experiments (ARPES) on cuprates like Nd$_{2-x}$Ce$_x$CuO$_4$ 
\cite{king,anders} appear to be consistent with LDA bandstructure 
calculations which, in turn, imply the validity of this theorem. 
Despite this flaw, the approximation (\ref{hubbardI}) yields a 
doping dependence for the Hall constant which is in good agreement 
with experiments on La$_{2-x}$Sr$_x$CuO$_4$. 

For another, it was pointed out in Refs.\ \cite{SSS} and \cite{SSS2},
that the high-frequency object $R_H^{\infty}$ is {\em not} directly
related to the location and topology of the Fermi surface. Instead, in
a strongly correlated system, the entire Brillouin zone tends to get
populated. In consequence, the weighted density average (\ref{a_d}) 
receives contributions from the entire Brillouin zone rather than just
from the vicinity of the Fermi surface. 

We can demonstrate this more explicitly by slightly changing the form
of the lower Hubbard band in Eq.\ (\ref{hubbardI}): We broaden the
delta function a little bit and shift some relatively small amount $Z$
of spectral weight to a new delta function contribution
$Z\delta(\omega-L(\epsilon_{\vec{k}}))$, with the function
$L(\epsilon_{\vec{k}})$ being undetermined yet. This amounts to
replacing Eq.\ (\ref{density}) by
$n_{\vec{k}\sigma}=h(\epsilon_{\vec{k}})$ with the function 
$h(\epsilon_{\vec{k}})$ differing from 
$\frac{1+\delta}{2}f(\frac{1+\delta}{2}\epsilon_{\vec{k}})$ only
in the following respects: The step at
$\epsilon_{\vec{k}}=\frac{2\epsilon_F}{1+\delta}$ and of height
$\frac{1+\delta}{2}$ is smoothed out while a new, much smaller one of
height $Z$ occurs at $\epsilon_{\vec{k}}=L^{-1}(\epsilon_F)$. This
last condition fixes the new Fermi surface. By choosing the function
$L(\epsilon)$ appropriately, we may place the Fermi surface wherever
we want. As long as $Z\ll\frac{1+\delta}{2}$, the crucial average
$a_d=\int_{-\infty}^{\infty}d\epsilon\,B(\epsilon)h(\epsilon)$ does
not differ very much from the result in Eq.\ (\ref{adad}). This
reasoning illustrates that, in the presence of strong correlations,
the doping dependence of $R_H^{\infty}$ is in fact uncorrelated to the
Fermi surface location. Also note, that it does not matter whether the
function $h(\epsilon)$ arises from coherent or incoherent excitations. 

\section{Hall constant in the low frequency regime}

In this section, we discuss the frequency dependence of the Hall
constant for frequencies {\em below} the Mott-Hubbard gap. Therefore, 
the appropriate model to start with is the $t-J$ model. It is 
straightforward to show that Eqs.\ (\ref{frha})-(\ref{Kofomega})
are still valid, however, with all quantities being redefined within 
the $t-J$ model \cite{fulde}. Apart from the redefinition of the
Liouville operator, this amounts to replacing the canonical Fermi
operators through projected ones in all quantities that appear, i.e.\  
$c_{i\sigma}\rightarrow X_i^{0\sigma}$, and 
$c_{i\sigma}^+\rightarrow X_i^{\sigma0}$. In particular, the current
operator is then given by Eq.\ (\ref{currentUinf}). If we renormalize 
the functions (\ref{SHofomega}) and (\ref{Kofomega}) according to
\begin{eqnarray}
 S_H(\omega;t-J)&=&-i<[\hat{J}_{x;0},\hat{J}_{y;0}]>s(\omega)\,,
\label{tJsofomega}\\
 K(\omega;t-J)&=&-i<[\hat{J}_{x;0},\hat{J}_{y;0}]>k(\omega)\,,
\label{tJkofomega}
\end{eqnarray}
the analog of Eq.\ (\ref{frha}) reads:
\begin{equation}
 R_H(z)=R_H^*\left(1+\int_{-\infty}^{\infty}d\omega\,
        k(\omega)\,\frac{\omega}{\omega-z}\right)\,.
\label{tJfrha}
\end{equation}
Here, $R_H^*$ is the infinite frequency Hall constant of the $t-J$
model that was already investigated in Ref.\ \cite{SSS}, and which has
been introduced in Eq.\ (\ref{result}) in the special case $J=0$. 
Furthermore, we have taken into account that the function $s(\omega)$ 
is normalized to unity, while $k(\omega)$ represents the unknown
memory function contribution. From the discussion in Sec.\
\ref{structureofSK}, we expect the function $k(\omega)$ to have only
one peak centered around zero frequency, because, in the $t-J$ model, 
doubly occupied sites can occur only virtually.

Before we set about discussing possibilities to calculate this
function via its moments, we try to gain some phenomenological
insight.  

\subsection{Phenomenological discussion \label{phenomenology}}

Very recently, the normal state ac-Hall constant was measured in 
YBa$_2$Cu$_3$O$_7$ thin films for frequencies up to 200 cm$^{-1}$
\cite{kaplan}. In this work, the experimental data have been fitted 
successfully in terms of parameters introduced by Anderson
\cite{anderson} to account for the observed $T^2$ dependence of the
inverse Hall angle in high-$T_c$ materials. Anderson's theory is 
based on spin charge separation with two different relaxation times 
and effective masses associated with the spinon and holon degrees of 
freedom: $\tau_{tr}$ is the decay time of the holons with effective 
mass $m_{tr}$, scattering off thermally excited spinons. On the other 
hand, a transverse relaxation rate $1/\tau_{H}$ is determined by the 
scattering between the spinons. Apart from this, $\sigma_{xx}$ and 
$\sigma_{xy}$ have the ordinary Drude form, i.e.\ 
$\sigma_{xx}\propto\tau_{tr}/m_{tr}$ and
$\sigma_{xy}=\sigma_{xx}\,\omega_c\tau_H$. Here, the cyclotron
motion is characterized by a mass $m_H$, i.e.\ $\omega_c\propto1/m_H$.
In Ref.\ \cite{kaplan}, Anderson's theory was extended to finite 
frequencies via the replacements
$\tau_{tr}\rightarrow\tau_{tr}/(1-i\omega\tau_{tr})$ 
and $\tau_{H}\rightarrow\tau_{H}/(1-i\omega\tau_{H})$. This led to the
following representation of the frequency dependent Hall constant:
\begin{equation}
 R_H(\omega)=\frac{m_{tr}}{m_H}\frac{1}{ne}
             \left(1+\frac{\tau_{H}-\tau_{tr}}{\tau_{tr}}
                     \frac{1}{1-i\omega\tau_{H}}\right)\,.
\label{phenfrha}
\end{equation}
This result is equivalent to the exact expression (\ref{tJfrha}), 
provided the following identifications are made:
\begin{eqnarray}
 k(\omega)&=&\frac{\tau_{H}-\tau_{tr}}{\tau_{tr}}\,
                           L_{1/\tau_{H}}(\omega)\,,
\label{kresphen}\\
 R_H^*&=&\frac{m_{tr}}{m_H}\frac{1}{ne}\,.
\end{eqnarray}
Here, $L_{\Gamma}(\omega)$ denotes the Lorentzian of width $\Gamma$ 
normalized to unity. Therefore, the unusual relaxation time $\tau_{H}$
is a measure for the width of the function (\ref{tJkofomega}) and thus
determines the decay rate of the correlation between the residual
forces $\hat{f}_{x;0}$ and $\hat{f}_{y;0}$ of the $t-J$ model (cf.\
Eqs.\ (\ref{residualforce}) and (\ref{Kofomega})). Furthermore, the
integrated weight of the function (\ref{kresphen}) measures the
deviation of the relaxation time $\tau_{H}$ from the ordinary
transport time $\tau_{tr}$. This deviation is related directly to that
of the Hall constant at zero frequency from its value at high
frequencies:
\begin{equation}
 \frac{\tau_{H}-\tau_{tr}}{\tau_{tr}}=\frac{R_H(\omega=0)-R_H^*}{R_H^*}\,.
\label{deviations}
\end{equation}
In the phenomenological expression (\ref{phenfrha}), the temperature
dependence is entirely contained in the two relaxation rates. They are
expected to vary as $\tau_{tr}\propto1/T$ and $\tau_{H}\propto1/T^2$ 
\cite{anderson}. For dimensional reasons, we take $\tau_{H}\propto
J/T^2$, since $J$ is the only energy scale characteristic of our
model. From these properties, we may infer the following: For one
thing, the memory function does not only describe the unusual
frequency dependence of the Hall constant in high-$T_c$
superconductors. Also, the observed anomalous temperature dependence
is mainly due to this memory function contribution. For another, we 
expect $\tau_{tr}$ to be relatively smaller than $\tau_{H}$ at low 
temperatures, since $\tau_H/\tau_{tr}\propto J/T$. Thus, Eq.\ 
(\ref{deviations}) suggests that the Hall constant {\em increases}
when zero frequency is approached. This enhancement was actually
observed in the above mentioned measurements on YBa$_2$Cu$_3$O$_{6+x}$
\cite{kaplan}. In any case, Eq.\ (\ref{deviations}) indicates that the
sign of the Hall constant is solely described by the high-frequency 
object $R_H^*$, as claimed in Ref.\ \cite{SSS}. 

Does the discussion so far point towards spin-charge separation as
advocated by Anderson? Obviously, the current operators
$\hat{J}_{\nu}$ are related to the charge degrees of
freedom only. Consequently, the spin physics must be accounted for by
the residual forces $\hat{f}_{\nu}$. This is also reflected by the 
proportionality $\tau_{H}\propto J/T^2$. However, up to now, we do not
have a compulsory argument why these residual forces should describe 
exclusively spin degrees of freedom.

In summary, Anderson's notion of two distinct relaxation rates is 
naturally backed up within the Mori theory. They may be interpreted as
the time scales set by the current operators and their associated 
residual forces.  

\subsection{Moments approach to the memory function}

Of course, it would be interesting to calculate the function
$k(\omega)$ of Eq.\ (\ref{tJfrha}) quantitatively within the $t-J$
model in order to relate its width and its integrated weight to the
parameters $t$, $J$, temperature $T$ and doping $\delta$. From 
the above discussion, we expect that the relevant information about 
its overall form may be put into only few parameters. As already 
mentioned, even two parameters as in Eq.\ (\ref{kresphen}) have been 
sufficient to obtain an excellent fit of experimental data 
\cite{kaplan}. In this subsection, we suggest the following procedure 
to construct $k(\omega)$: Parametrize this function by $n$ parameters
which are subsequently fixed by its first $n$ moments. Up to now, this
seems to be the only reliable way to take into account the 
superprojectors $Q$. Of course, all moments  
\begin{equation}
 k_l=\int_{-\infty}^{\infty}d\omega\,k(\omega)\omega^{2l}\,
\label{momk}
\end{equation}
exist, why we have to replace the Lorentzian (\ref{kresphen}) by a 
``short-range'' function, e.g.\ a Gaussian multiplied by a
polynomial. We proceed by relating the moments of $k(\omega)$ to that 
of the function $s(\omega)$ and the optical conductivity. Finally, we 
discuss a possibility to calculate these moments. 

The ordinary conductivity is given in Mori theory by the following 
expression \cite{lange}: 
\begin{equation}
 \sigma_{xx}(\omega+i0^+)=
   ie^2\beta(\hat{J}_{x;0}|\frac{1}{\omega+L+i0^+}|\hat{J}_{x;0})\,,
\end{equation}
the real part of which may be shown to be an even function of
$\omega$ \cite{fick}, and, due to a sum rule, can be written in terms 
of a function $c(\omega)$, that is normalized to unity:
\begin{equation}
 \Re\{\sigma_{xx}(\omega+i0^+)\}=\pi e^2\chi^0c(\omega)\,.
\label{optcond}
\end{equation}
In the appendix, we show that all the moments (\ref{momk}) may be
traced back recursively to that of the functions (\ref{tJsofomega})
and (\ref{optcond}):
\begin{eqnarray}
 s_l&=&\int_{-\infty}^{\infty}d\omega\,s(\omega)\omega^{2l}\,,
\label{moms}\\
 c_l&=&\int_{-\infty}^{\infty}d\omega\,c(\omega)\omega^{2l}\,,
\label{momc}
\end{eqnarray}
provided $l\ge1$. Thereby, only the definition of the superprojectors 
$Q$ has to be used in Eq.\ (\ref{Kmom}). The result may be written as 
follows:
\begin{equation}
 k_l=-s_l+\sum_{j=1}^{l}a_js_{l-j}\quad\mbox{for }l\ge1\,,
\label{momkres}
\end{equation}
where the coefficients $a_j$ are polynomials in the moments
(\ref{momc}) and are listed in the appendix up to $j=6$. If we had a
good method for calculating the moments (\ref{moms}) and (\ref{momc}),
we could construct the unknown function $k(\omega)$ via its first $n$ 
moments as already explained. This approximation is reliable, if the 
zeroth moment $k_0$, calculated from Eq.\ (\ref{momk}), converges fast
enough with increasing $n$. Since in the $t-J$ model, $k(\omega)$ has 
only one peak around $\omega=0$, the first few moments are expected to
be sufficient for this to happen. An advantage of this approach is 
that moments are {\em global} properties of a spectral function and 
as such are less sensitive to its detailed resonance structure and
to approximations involved. Also, some approximation schemes are
better suited for the calculation of moments rather than the
underlying spectral function. 

For instance, within a high-temperature expansion, moments are
accessible, at least in principle, while the corresponding spectral 
function is not. However, in the case of the moments (\ref{moms}),
only very few moments, and to only low orders in $1/T$, are within
reach. This is all the worse, since now, we are interested in the Hall
constant at low frequencies, i.e.\ no other high-energy scale is
present as it was in the context of $R_H^*$. 

Another example is the exact diagonalization technique \cite{dagotto}. 
In this method, spectral functions are calculated numerically via the 
exact eigenstates and eigenenergies on the basis of their Lehmann
representation. While being exact, the intrinsic problem of this
method is the constraint of working on relatively small clusters. 
Therefore, the delta functions have to be broadened in order to obtain 
smooth functions. In contrast to this, no additional approximation is
required when calculating moments. Of course, the smallness of the 
clusters remains the major restriction of this method. Nevertheless,
when combined with the moments approach as suggested above, it is a 
means of extracting reliable information about the frequency dependent
Hall constant within the $t-J$ model and should therefore be the
subject of a future work. 

\section{Discussion and Conclusions}

In this paper, we have derived first results for the Hall effect in 
correlated electron systems in the strong correlation limit within the
recently developed memory function approach \cite{lange}. We
have focused our attention mainly on the memory function term, which
is neglected in Boltzmann type approaches. The important new physics
to be expected from this contribution comprises the unusual frequency 
{\em and} temperature dependence of the Hall constant, observed in
the normal state of high-$T_c$ superconductors. 

In the single-band Hubbard model, a finite amount of spectral weight
for particle-hole excitations, caused by the Hall current, is always 
pinned at the energy $U$. This is valid no matter how large the
correlation strength $U$ is. We have shown that the memory function
removes these high-energy excitations in the limit
$U\rightarrow\infty$, thus accounting for the frequency dependence of 
the Hall constant down to frequencies within the charge transfer gap. 
The corresponding decrease of the Hall constant by a factor 
$(1+\delta)/2$ was calculated exactly to leading order in $1/T$ and 
corroborated within an approximate treatment, valid at $T=0$. However,
our analysis did not provide us with information about the frequency 
dependence of the Hall constant at lower frequencies. The reason is
that it was based on moments, which are dominated by the
high-frequency contributions. We have also calculated the infinite 
frequency Hall constant analytically within the so-called Hubbard I 
approximation. In essence, we recovered the exact result for 
$U=\infty$ and $T\rightarrow\infty$ and explained, why this result
does not rely on the location of the Fermi surface. 

Finally, the Hall constant at low frequencies was investigated within 
the $t-J$ model, an effective model acting in the reduced Hilbert
space without doubly occupied sites. We observed that our memory 
function formalism distinguishes inherently between two time scales: 
Firstly, the dynamics of the current operators is characterized by the
ordinary transport relaxation time $\tau_{tr}$. And secondly, the 
impact of all the other degrees of freedom on this charge transport is
taken into account by fluctuating forces, that introduce an unusual 
time scale $\tau_{H}$. On the other hand, it was pointed out by 
Anderson, that temperature dependences of transport and Hall effect 
measurements can best be understood in terms of two relaxation times, 
following a $1/T$ and $1/T^2$ law \cite{anderson}. We have shown that
the time scales encountered within the Mori theory are identical to 
those introduced by Anderson. Furthermore, we have shown that the 
deviation of the unusual decay time $\tau_{H}$ from the ordinary 
transport time $\tau_{tr}$ is intimately connected to the frequency 
dependence of the Hall constant at low frequencies. Thus, the 
temperature and frequency dependence of the Hall constant result from 
each other and are both due to the memory function contribution. It
would be very interesting to investigate this interplay further, both
theoretically and experimentally. As for the theoretical side, we have
proposed an approach based on moments. It allows the exact treatment
of the superprojector that reflects the distinction between the two 
time scales, leaving us with the problem of finding the first few 
moments of the ordinary current-current correlation functions. Except 
for a well studied prefactor, the memory function term is mainly 
determined by two parameters. Therefore, we expect its first few 
moments to provide us with enough information to fix these
parameters. 

\acknowledgments
The author is indebted to P.\ W\"olfle for many stimulating
discussions. This work has been supported by the
Landesforschungsschwerpunktprogramm and the Sonderforschungsbereich
195. 

\begin{appendix}

\section{Reduction of the moments of the memory function}
 
First of all, we recall the definition of the moments (\ref{Smomdef})
and (\ref{Kmomdef}), however with all operators and superoperators now
being redefined within the $t-J$ model. They are related to the
moments (\ref{momk}) and (\ref{moms}) via the equations 
$K_l=-i<[\hat{J}_{x;0},\hat{J}_{y;0}]>k_l$ and
$S_l=-i<[\hat{J}_{x;0},\hat{J}_{y;0}]>s_l$. In addition, we require the 
moments (\ref{momc}), renormalized as $C_l=\chi^0c_l$. Next, we define
the quantities
\begin{eqnarray}
 X_l^n&:=&i\beta(L^n\hat{J}_{x;0}|\underbrace{LQ\;\ldots\;QL}_{\mbox{$l$
 operators $L$}}|\hat{J}_{y;0})\,,
\label{Xln_app}\\
 Y_l^n&:=&\beta(L^n\hat{J}_{x;0}|\underbrace{LQ\;\ldots\;QL}_{\mbox{$l$
 operators $L$}}|\hat{J}_{x;0})
\label{Yln_app}\,,
\end{eqnarray}
for $l\ge1$. By using the definition (\ref{projectorQ}) for each of
the first superprojector on the right hand sides, we obtain the
following recursion relations:
\begin{eqnarray}
 X_l^n&=&X_{l-1}^{n+1}-\frac{Y_1^n}{\chi^0}X_{l-1}^0
                            \quad\mbox{for $n$ odd}\,,
\label{recrel1}\\
 X_l^n&=&X_{l-1}^{n+1}-\frac{X_1^n}{\chi^0}Y_{l-1}^0
                            \quad\mbox{for $n$ even}\,,\\
 Y_l^n&=&Y_{l-1}^{n+1}-\frac{Y_1^n}{\chi^0}Y_{l-1}^0
                            \quad\mbox{for $n$ odd}\,,\\
 Y_l^n&=&Y_{l-1}^{n+1}\quad\mbox{for $n$ even}\,.
\label{recrel4}
\end{eqnarray}
To prove these relations, we only have to use the following facts:
$X_1^n$ and $Y_1^n$ vanishes for all odd and even integers $n$,
respectively. This is due to the fact that the functions
(\ref{SHofomega}) and (\ref{optcond}) are even, hence their odd
moments vanish. Moreover, the quantities (\ref{Xln_app}) are of first
order in the magnetic field. To relate the unknown moments $K_l$ to
the moments $S_l$ and $C_l$, we have to supplement the recursion
formulas (\ref{recrel1})-(\ref{recrel4}) by the following equations:
\begin{eqnarray}
 K_l&=&X_{2l+1}^0\,,\\
 S_l&=&-X_1^{2l}\,,\\
 C_l&=&Y_1^{2l-1}\,.
\end{eqnarray}
However, only even numbers of iterations occur. Therefore, we may
combine two successive iteration steps into one. This leads to the
following effective recursion relations:
\begin{eqnarray}
 H_l^n&=&H_{l-1}^{n+1}-\frac{C_{n+1}}{C_0}H_{l-1}^0
                      +\frac{S_n}{C_0}N_{l-1}^1\,,\\
 N_l^n&=&N_{l-1}^{n+1}-\frac{C_n}{C_0}N_{l-1}^1\,,
\end{eqnarray}
where we have defined
\begin{eqnarray}
 H_l^n&:=&X_{2l+1}^{2n}\qquad\;\mbox{for $n\ge0$, $l\ge0$}\,,\\
 N_l^n&:=&Y_{2l+1}^{2n-1}\qquad\mbox{for $n\ge1$, $l\ge0$}\,,
\end{eqnarray}
and where the contact to the moments $K_l$, $S_l$ and $C_l$ is
established by means of the equations
\begin{eqnarray}
 K_l&=&H_l^0\,,\\
 S_l&=&-H_0^l\,,\\
 C_l&=&N_0^l\,.
\end{eqnarray}
This recursive procedure results in Eq.\ (\ref{momkres}) with the
first six coefficients being given as follows:
\begin{eqnarray}
 a_1&=&2c_1\,,\\
 a_2&=&2c_2-3c_1^2\,,\\
 a_3&=&2c_3-6c_1c_2+4c_1^3\,,\\
 a_4&=&2c_4-6c_1c_3-3c_2^2+12c_1^2c_2-5c_1^4\,,\\
 a_5&=&2c_5-6c_1c_4-6c_2c_3+12c_1^2c_3+12c_1c_2^2\nonumber\\
    &-&20c_1^3c_2+6c_1^5\,,\\
 a_6&=&2c_6-6c_1c_5-6c_2c_4+12c_1^2c_4-3c_3^2\nonumber\\
    &+&24c_1c_2c_3-20c_1^3c_3+4c_2^3-30c_1^2c_2^2\nonumber\\
    &+&30c_1^4c_2-7c_1^6\,.
\end{eqnarray} 

\end{appendix}

\end{multicols}


\begin{references}

\bibitem{lange}E. Lange, cond-mat/9606176, to appear in Phys. Rev. B
together with the present work.
\bibitem{ong1}N. P. Ong, Physical Properties of High Temperature 
Superconductors, edited by D. M. Ginsberg, World Scientific, 
Singapore (1990), Vol. 2.
\bibitem{anderson}P. W. Anderson, Phys. Rev. Lett. {\bf 67}, 2092
(1991). 
\bibitem{peierls}L. Friedman, T. Holstein, Ann. Phys. (N. Y.) {\bf
21}, 494 (1963); Phys. Rev. {\bf 165}, 1019 (1968); K. G. Wilson,
Phys. Rev. D {\bf 10}, 2445 (1974).
\bibitem{fick}E. Fick, G. Sauermann, The Quantum Statistics of 
Dynamic Processes, Springer (1990).
\bibitem{SSS}B. S. Shastry, B. I. Shraiman, R. P. Singh, Phys. Rev. 
Lett. {\bf 70}, 2004 (1993).  
\bibitem{bacci}S. Bacci, E. Gagliano, R. Martin, J. Annett,
Phys. Rev. B {\bf 44}, 7504 (1991). 
\bibitem{harrislange}B. Harris, R. Lange, Phys. Rev. {\bf 157}, 295
(1967). 
\bibitem{donald}A. H. MacDonald, S. M. Girvin, D. Yoshioka,
Phys. Rev. B {\bf 37}, 9753 (1988).
\bibitem{eskes}H. Eskes, A. Oles, M. Meinders, W. Stephan,
Phys. Rev. B {\bf 50}, 17980 (1994).
\bibitem{thompson}C. J. Thompson et al., J. Phys. A: Math. Ge. {\bf
24}, 1261 (1991).
\bibitem{voruganti}P. Voruganti, A. Golubentsev, S. John, Phys. Rev. B
{\bf 45}, 13945 (1992). 
\bibitem{fetter}A. L. Fetter, J. D. Walecka, Quantum Theory of
Many-Particle Systems, McGraw-Hill (1971).
\bibitem{hubIapprox}J. Hubbard, Proc. R. Soc. London, Ser. A {\bf
276}, 238 (1963). 
\bibitem{fukcomment}In Ref.\ \cite{fukuyama}, the Hall effect was
treated numerically within the Hubbard I approximation on the basis of
the Boltzmann equation. This amounts to the same thing as considering 
$R_H^{\infty}$ \cite{lange}.
\bibitem{fukuyama}H. Fukuyama, Y. Hasegawa, Physica B {\bf 148}, 204
(1987). 
\bibitem{agd}A. Abrikosov, L. P. Gorkov, I. E. Dzyaloshinskii, Methods
of Quantum Field Theory in Statistical Physics, Pergamon Press (1965).
\bibitem{king}D. M. King et al., Phys. Rev. Lett. {\bf 70}, 3159
(1993). 
\bibitem{anders}R. O. Anderson et al., Phys. Rev. Lett. {\bf 70}, 3163
(1993). 
\bibitem{SSS2}B. S. Shastry, B. I. Shraiman, R. P. Singh, Phys. Rev. 
Lett. {\bf 71}, 2838 (1993).  
\bibitem{fulde}P. Fulde: Electron Correlations in Molecules and
Solids. Springer, 1991.
\bibitem{kaplan}S. G. Kaplan et al., Phys. Rev. Lett. {\bf 76}, 696
(1996). 
\bibitem{dagotto}For a review, see E. Dagotto, Rev. Mod. Phys., {\bf
66}, 763 (1994).

\end{references}
\end{document}